\documentclass[referee,a4paper,12pt,traditabstract]{jswsc} 

\usepackage{graphicx}
\usepackage{txfonts}
\usepackage{subfigure}
\usepackage{epstopdf}
\usepackage[displaymath,mathlines]{lineno}
\usepackage[authoryear,round]{natbib}
\usepackage[backref]{hyperref}
\usepackage{url}

\hypersetup{colorlinks=true,citecolor=cyan,urlcolor=cyan,linkcolor=blue}

\begin{document}


   \title{Statistics of Large Impulsive Magnetic Events In The Auroral Zone}
   
   \authorrunning{Reiter et al.}
   \titlerunning{Statistics of Large Impulsive Magnetic Events in The Auroral Zone}

    \author{K. Reiter\inst{1}\thanks{Corresponding author: \href{mailto:kyle.reiter1@ucalgary.ca}{kyle.reiter1@ucalgary.ca}}
        \and S. Guillon\inst{2}
        \and M. Connors\inst{3,1}
        \and B. Jackel\inst{1}}

    \institute{Department of Physics and Astronomy, University of Calgary, Calgary, Canada
            \and Trans\'{E}nergie-\'{E}quipement, Hydro-Qu{\'{e}}bec, Montr\'eal, Canada
            \and Athabasca University Observatories, Athabasca, Canada
            }


 
  \abstract
   {Impulsive geomagnetic variations, latitudinally localized to the auroral zone, have been observed over the period from 2015-2020. These impulsive events have been observed mostly in the pre-midnight sector as upward vertical perturbations. Diurnal variations in geomagnetically-triggered harmonic distortion events observed in Hydro-Qu\'ebec's Syst\`eme de Mesure de D\'ecalage Angulaire (SMDA) synchrophasor measurement system have been found to have a peak in the number of events around midnight. This was similar to diurnal rates of occurrence of negative $B_z$ geomagnetic impulsive events, observed at nearby auroral zone magnetometers. Superposed epoch analysis demonstrates the impulses are regularly associated with increases in harmonic distortion observed at a nearby substation transformer. These large impulsive vertical geomagnetic perturbations appear to be local vortical ionospheric disturbances.}


   \maketitle

\section{Introduction}
The nightside auroral zone ionosphere has long been a subject of study \citep{Fukushima1994}, being a region of dynamic current structures which interconnect with the inner magnetosphere \citep{McPherron1973}. During geomagnetically active periods, changing ionospheric currents occur in the auroral zone. Resulting geoelectric fields present a potential hazard: the excitation of geomagnetically induced currents (GIC) in conducting infrastructure such as power transmission networks and pipelines \citep{Bolduc2002,Boteler2003,Lundstedt2006}. The relatively low frequency of GIC variations, compared to the operating frequency of the power transmission network, means GIC can produce a quasi-DC voltage bias of transformers. This can saturate the core of the transformer, producing harmonic distortion of the transformer voltage \citep{Boteler2001}. For auroral zone power infrastructure, substorms, pulsations, and horizontal geomagnetic perturbations constitute major drivers of geomagnetic activity which may cause power system disruption \citep{Guillon2016,Boteler2019,Kappenman2005,Pulkkinen2003,Stauning2013,Ngwira2015}. It is possible that large impulsive magnetic events may act as another source of GIC. If these events were to have a prominent vertical component, Faraday's Law would indicate that the resultant large vertical magnetic flux changes produced from these events should induce similarly large horizontal geoelectric fields. Commonly, when modeling geoelectric fields based on ground magnetometer measurements, $B_z$ measurements are neglected, using the plane-wave assumption \citep{WaitJamesRG,Boteler2017}. We present a statistical study including the occurrence of $B_z$ impulsive events and their relation to GIC.

Harmonic distortion (HD) is often used as an indicator of the stress applied to a power transmission network due to GIC \citep{NZHD,Clilverd2020,Bolduc2002}. GIC in transformers within a power transmission network have a spectral content with frequencies much lower than the normal 60 Hz operating frequency of the network. Therefore they can be approximated as quasi-DC currents. Transformer cores within a network may become partially saturated due to the introduction of GIC, resulting in distortion of the AC signal. This can be quantified as HD \citep{GICHD}. The Hydro-Qu\'ebec network has been extensively instrumented to measure HD, which provides a continuous record of space weather impacts to the network \citep{Kamwa2006}. As a result, this study uses HD as an indicator of GIC-driven stress on the power transmission network.

Previous statistical studies of impulsive geomagnetic events examined impulses measured on the dayside in the cusp regions \citep{doi:10.1029/GL013i011p01089,Lanzerotti1991}. One study considered auroral-zone impulsive events but limited to horizontal geomagnetic perturbations \citep{Moretto2004}. Some more recent studies have looked at impulsive geomagnetic perturbations on the nightside, but have principally focused on impulsive perturbations in the polar cap \citep{Engebretson2019} or in a storm context \citep{Engebretson2019b}. This study seeks to characterize the latitudinal distribution of impulsive magnetic events, their diurnal distribution in magnetic local time (MLT), their amplitude, which component is dominant, and their relation to power network disturbances. We characterize these events in the aggregate and find that most geoeffective events are associated with vertical disturbances.

\section{Materials and Methods}
\subsection{Source Data}

\begin{figure}[ht]
\includegraphics[width=35pc]{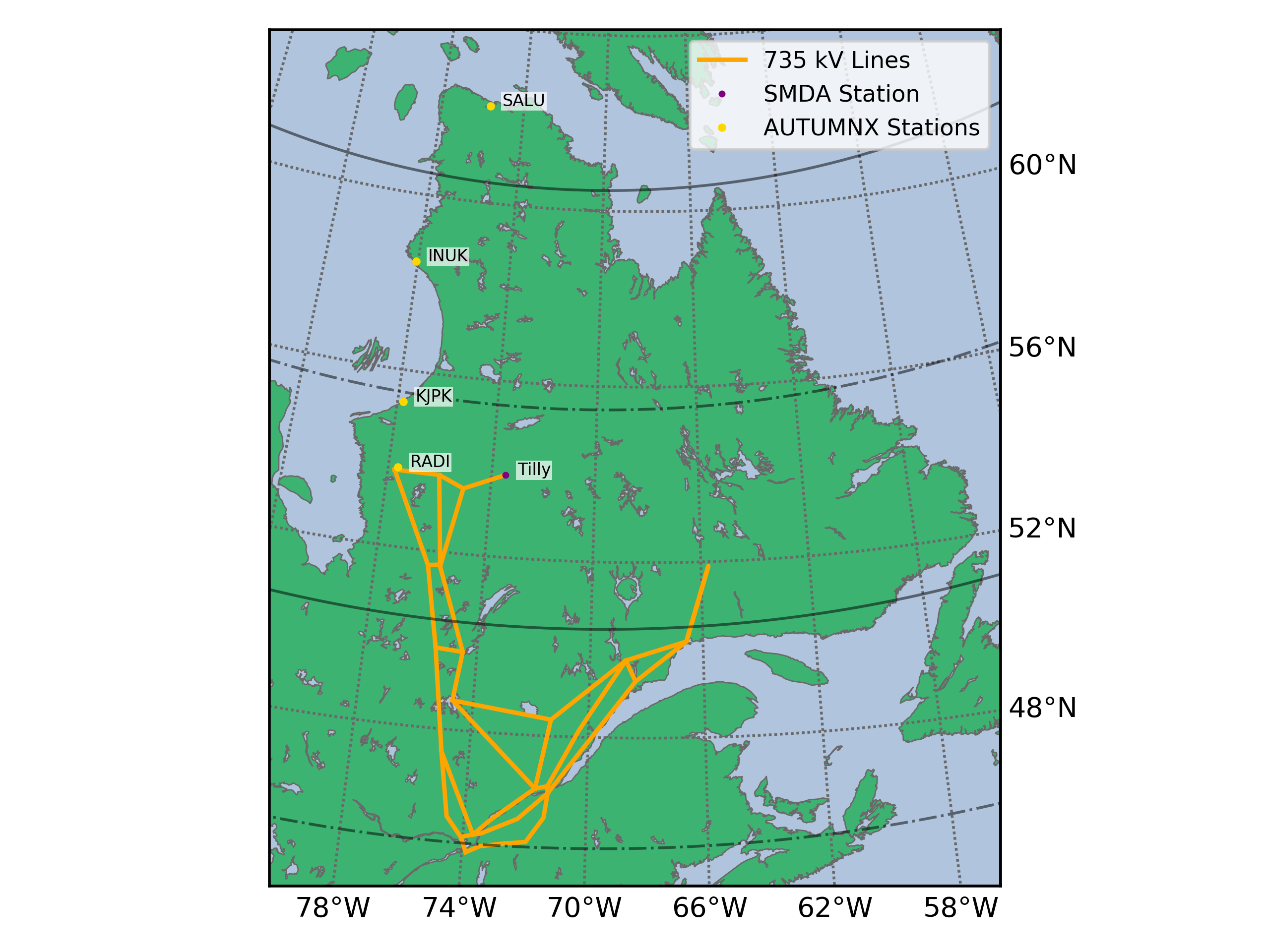}
\caption{The region of interest. Magnetometers used are indicated in yellow. 735 kV Hydro-Qu\'ebec transmission lines are indicated in orange, with  the Tilly transformer substation from which EHD data is used indicated in purple. Dot-dash lines indicate 55$^\circ$ and 65$^\circ$ corrected geomagnetic latitude, solid lines indicate 60$^\circ$ and 70$^\circ$ latitude.}
\label{map}
\end{figure}

We analyze data from two sources: ground-based magnetometers and synchrophasor measurements of voltage distortion at substation transformers. The region of interest for the study is shown in Figure \ref{map}. The fluxgate magnetometers used in this study were of the same design used by Time History of Events and Macroscale Interactions during Substorms (THEMIS) ground-based observatories (GBO) \citep{Harris2009}. Magnetic field data is from selected stations in the Athabasca University THEMIS UCLA Magnetometer Network eXpansion (AUTUMNX) \citep{AUTUMNX} ranging from subauroral to polar cap regions. AUTUMNX stations in this study were Radisson (RADI, 63.25$^\circ$ geomagnetic latitude), Kuujjuarapik (KJPK, 64.73$^\circ$), Inukjuak (INUK, 67.92$^\circ$), and Salluit (SALU, 71.71$^\circ$). Corrected Geomagnetic latitudes for these stations were determined by \citet{AUTUMNX}. The magnetometers are located along a longitudinal chain at an MLT where midnight is at 5 UT. Magnetometer data was collected and stored at a 2 Hz cadence, in north (X) west to east (Y), and positive downward vertical (Z) components. Data from the AUTUMNX network has been analyzed from January 1, 2015 through  June 30, 2020.

Hydro-Qu\'ebec total harmonic distortion data from June 2014 through May 2017 was collected on a 4 s cadence from automatically selected periods of elevated disturbance identified by Hydro-Qu\'ebec. This is measured using Hydro-Qu\'ebec's Syst\`eme de Mesure de D\'ecalage Angulaire (SMDA) synchrophasor measurement system \citep{Guillon2016,Kamwa2006}. Total harmonic distortion is a measure of the voltage distortion at the substation transformer. These are archived when even-harmonic distortion (EHD) of voltages at designated substation transformers throughout the Hydro-Qu\'ebec network exceeds substation-specific thresholds. This triggers the recording of high-resolution HD data for a period of approximately 30 minutes. The start time of SMDA data collection serves as the event time used. EHD was collected at a 1 Hz cadence, and is defined by Hydro-Qu\'ebec as:

\begin{linenomath*}
 \begin{equation}
 EHD = \frac{\sqrt{V_2^2+V_4^2+V_6^2+V_8^2}}{\sqrt{V_{f}^2+V_2^2+V_4^2+V_6^2+V_8^2}}
 \label{EHD}
 \end{equation}
\end{linenomath*}
 
 $V_n$ corresponds to the amplitude of the $n$th harmonic of voltage at the fundamental 60 Hz operating frequency, $V_f$. Odd harmonic distortion information was not used in the analysis due to a lack of availability. While the high-resolution HD data is available only on a limited and activity-triggered basis, lower resolution EHD data is available on a continuous basis. The EHD data is a continuous time series from 2015-2019, collected from the Tilly substation (about 5$^\circ$ south of Inukjuak) as seen in Figure \ref{map}.

\subsection{Geomagnetic Extremum Identification}
\label{extremum}

\begin{figure}[ht]
\centering
\includegraphics[width=\textwidth]{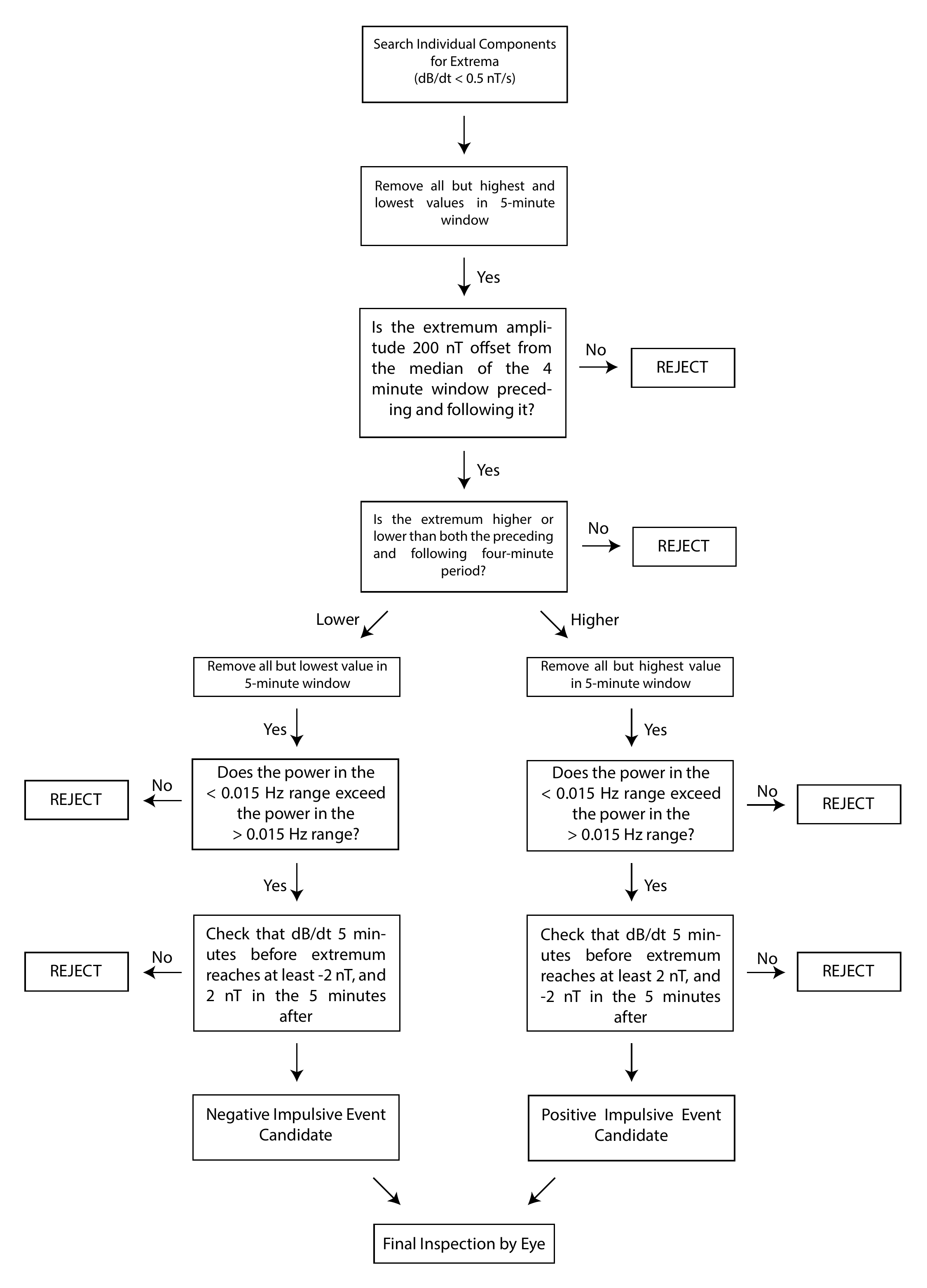}
\caption{A flow diagram depicting the selection criteria for impulsive events. Selection is conducted on a component-by-component basis for each individual magnetometer.}
\label{flowsel}
\end{figure}

We designed an algorithm to select for geomagnetic perturbations which were short in duration (minutes), large in amplitude (100s of nT), and non-periodic. This led to a set of criteria similar to those used by \citet{Lanzerotti1991}, as detailed in Figure \ref{flowsel}. While large impulsive magnetic events are associated with large $dB/dt$, these are distinguished from monotonic decreases or increases in the field by an extremum. $dB/dt$ calculated from field measurements at the extremum will be small in magnitude, reflecting the flat-topped characteristic of impulsive events. As an initial selection criterion to identify local extrema, a low $dB/dt$ threshold was used. Subsequent removal of all but the largest and smallest values from five-minute windows in the data set was used to identify the largest magnitude impulsive events in the window. A 200 nT offset of the median values of 4-minute windows on both sides of the extrema was chosen. This selects for large perturbations which may result in the induction of significant GIC. Impulsive events are sorted into positive or negative events by their relation to the median component values in the four minutes preceding and following the extremum. Examples of events which do and do not meet the selection criteria can be found in Annex A.

The power spectral density was determined from a two-hour window (14400 points) around the event, which is mean-subtracted with a Hanning (weighted cosine) window applied \citep{KanasewichE.R1981Tsai}: 

\begin{linenomath*}
 \begin{equation}
w(n)=0.5-0.5\cos\Big(\frac{2\pi n}{M-1}\Big)\;\;\;\;\;\; 0\leq n\leq M-1
 \end{equation}
\end{linenomath*}

$w(n)$ defines the windowing function which the time-series data is multiplied element-wise by, $n$ is the index number, and $M$ is the size of the window. Events with missing data were not used in the analysis. For each event, the power spectrum was calculated using a discrete Fourier transform implemented using a fast Fourier transform algorithm. 

The power content in the $<$ 1.5 mHz range was compared to the 1.5 - 16.7 mHz range, and events where more power was contained in the higher frequency range were discarded. This principally removes pulsation events with periodicities similar to the 8-minute selection window, as well as candidate events which are the result of instrumentation issues. The last step in computer-identification of events was to check that the $dB/dt$ in the 10 minute window immediately preceding and following the candidate event has the correct magnitude and sign for an impulsive event of its identified polarity. This removes events which may have local component extrema but are not characterized by sudden changes. Finally, candidates are examined by eye to remove any events characteristic of instrumentation faults. This step also removes any events or variations which are insignificant when compared with proximate geomagnetic activity. This step results in a relatively small number of candidate events being rejected: in the case of the Inukjuak magnetometer, only 69 of 996 total events. Similar human interventions have been deemed necessary in previous impulsive event statistical studies \citep{Lanzerotti1991}.

While the selection algorithm takes inspiration from \citet{Lanzerotti1991}, the selection criteria have been modified so as to select for similar but distinct types of magnetic perturbations. Principal differences include the larger and more selective magnetic component deflection threshold, and the use of spectral power content as a filter. Most notably, our analysis allows for events occurring in a single geomagnetic field component. This broadens our analysis as compared to that of \citet{Lanzerotti1991}, allowing for comparison of the relative prevalence of horizontal and vertical impulsive events.

\section{Analysis}
\subsection{Diurnal Variations}

\begin{figure}[ht]
\centering
\includegraphics[width=30pc]{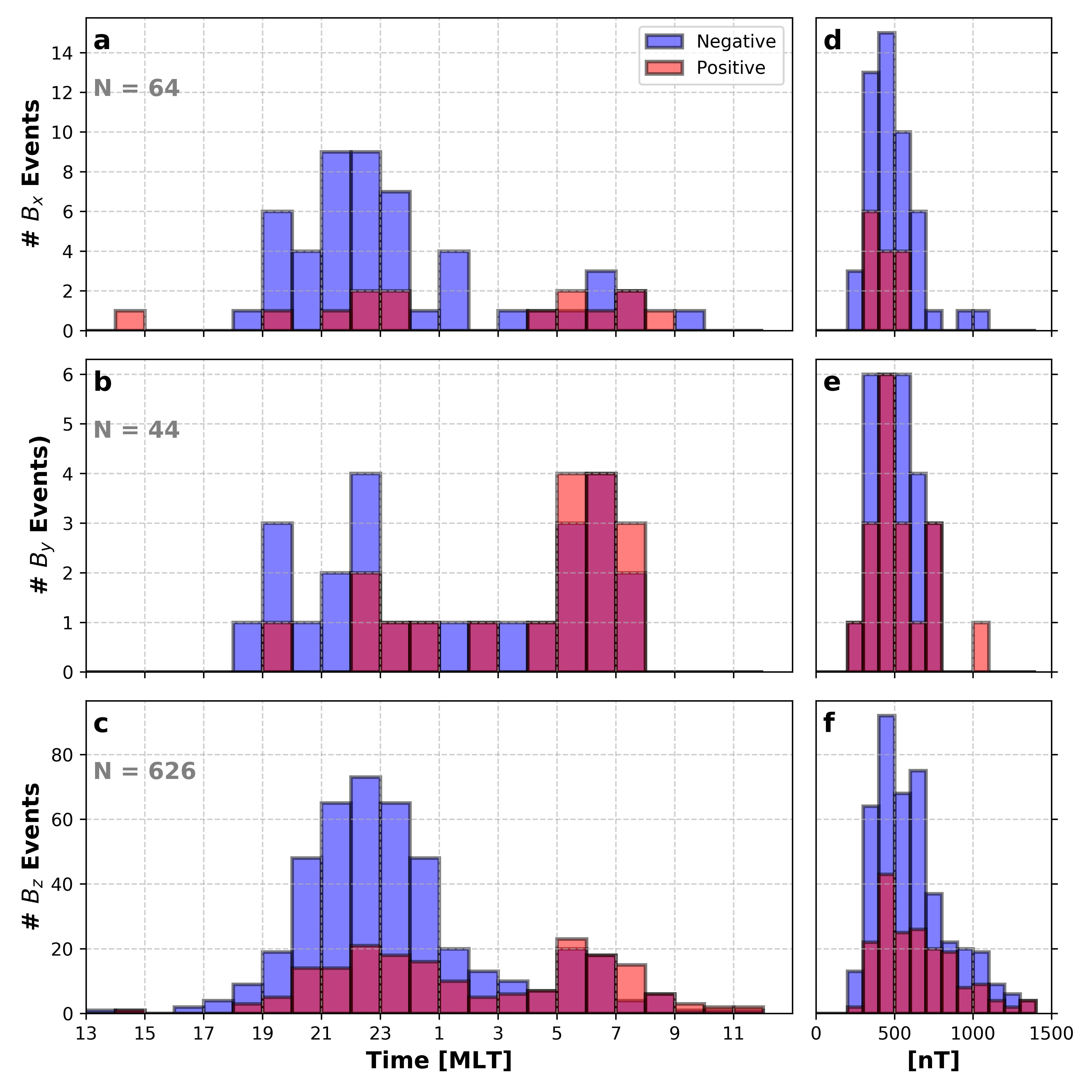}
\caption{Left (a-c): Diurnal variations in impulsive magnetic events per hour found from 2015-2020 in Inukjuak, Quebec. Right (d-f): Distribution of magnitude of 8-minute component ranges around identified impulsive events per 100 nT bin. Vertical scales vary between components to better convey diurnal trends. Total event numbers are indicated in bold gray for each component. Blue shading indicates negative impulses, red shading positive.}
\label{todplotINUK}
\end{figure}

\begin{figure}[ht]
\centering
\includegraphics[width=30pc]{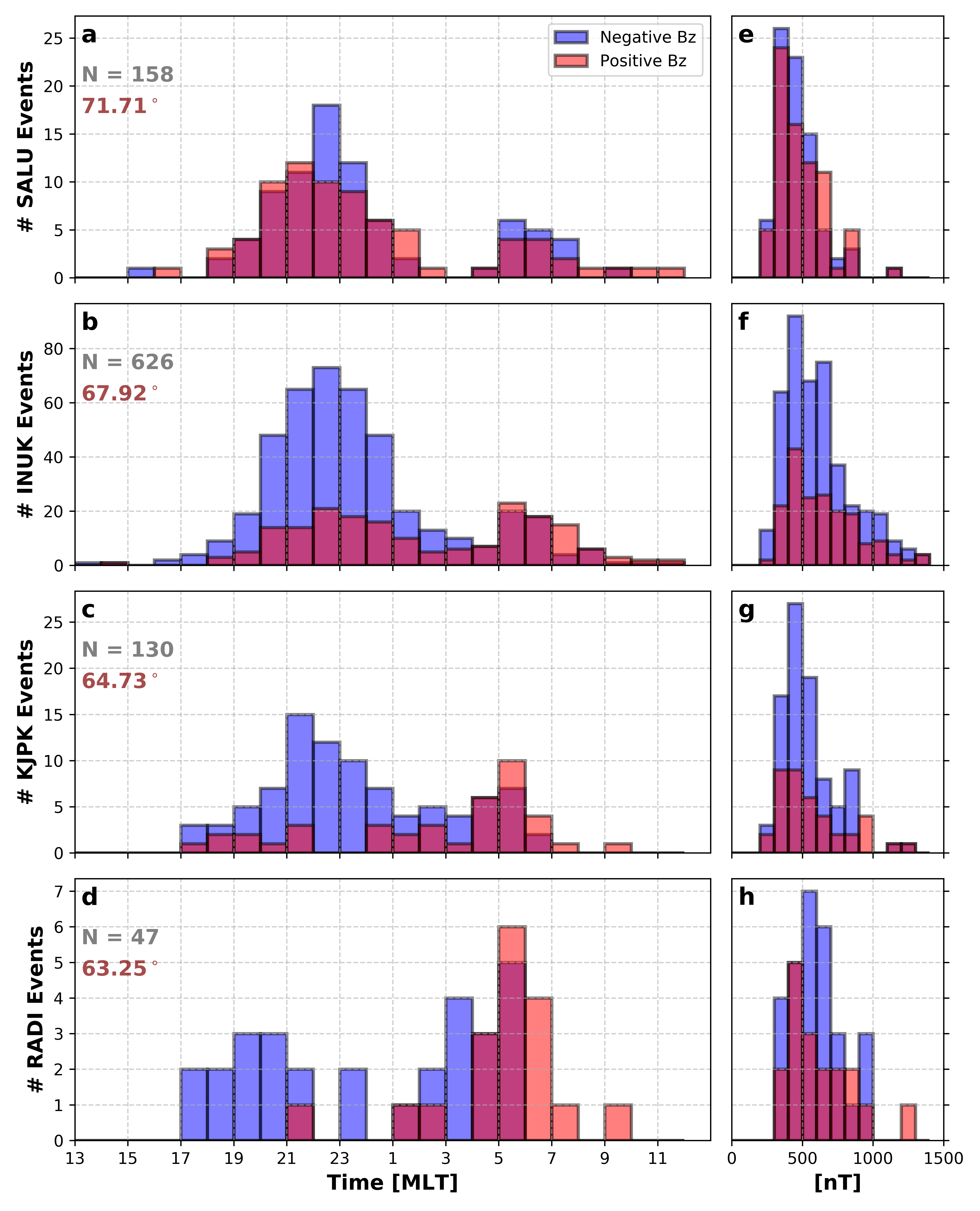}
\caption{Left (a-d): Magnetic local time distribution of impulsive events per hour as recorded at selected AUTUMNX locations. Right (e-h): Distribution of magnitude of 8-minute $B_z$ ranges around identified impulsive events per 100 nT bin. Vertical scales vary between magnetometers to better convey diurnal trends. Total event numbers are indicated in bold gray for each magnetometer, with corrected geomagnetic latitude indicated in dark red indicated underneath. Blue shading indicates negative impulses, red shading positive.}
\label{todplots}
\end{figure}

Patterns in the occurrence of impulsive events were identified by constructing histograms using event magnitude and MLT hourly bins. Diurnal variations and the magnitude distribution of magnetic impulsive events for the field components at Inukjuak, Qu\'ebec are seen in Figure \ref{todplotINUK}. This auroral-zone magnetometer shows a distinct peak in occurrence of impulses in all components in the pre-midnight and morning sectors. The rate of occurrence and magnitude of $B_z$ perturbations are significantly higher than those of other components. The diurnal distribution of impulsive events for magnetometer sites from sub-auroral through polar-cap latitudes can be seen in Figure \ref{todplots}. The total number of $B_z$ events for each magnetometer can be seen in gray in panels a-d. High rates of occurrence of impulsive $B_z$ perturbations are narrowly latitudinally localized. As seen in panel f of Figure \ref{todplots}, an extended tail of large magnitude impulsive events is evident at Inukjuak, which also experienced by far the most $B_z$ impulsive events. The magnitude of these impulses is significantly larger than those in similar previous studies \citep{Lanzerotti1991,Moretto2004}. The largest impulsive events see an 8-minute range in excess of 1000 nT in the vertical component. The 8-minute range is defined analogously to the hourly range of one geomagnetic component, for one magnetometer \citep{Hruska1987,Danskin2015}. Perturbations in multiple components were seen for many events, though usually not of a magnitude to qualify as simultaneous impulsive events in multiple components. For example, only 38 of 436 negative $B_z$ impulsive events at Inukjuak had coincident impulsive events in another component.

\begin{figure}[ht]
\centering
\includegraphics[width=30pc]{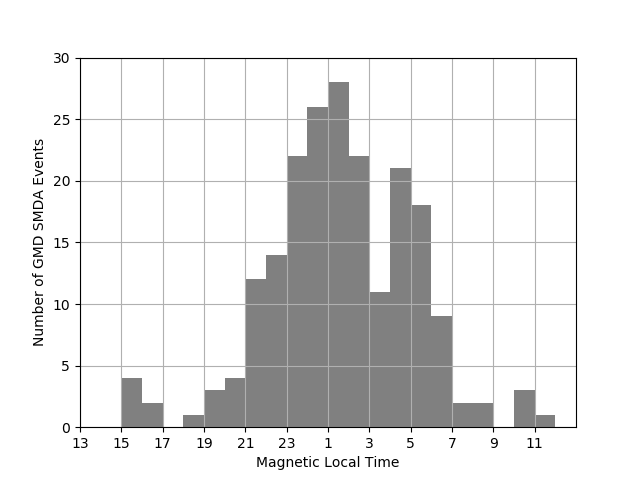}
\caption{Local time distribution of harmonic distortion events observed by Hydro-Qu\'ebec SMDA monitoring system, 2014-2017. Local time was taken from an average of longitudinal extremes of substations which may trigger the event.}
\label{SMDA-start}
\end{figure}

It is of interest to power network operators to know at what local times heightened levels of HD are prevalent in their networks. The diurnal distribution of heightened HD within a power transmission network may also provide insight into potential drivers of HD and related GIC. Figure \ref{SMDA-start} shows the local-time distribution of EHD threshold triggered HD events in the Hydro-Qu\'ebec network, which occur predominantly on the nightside. Peaks are seen around midnight and in the morning sector around 5 MLT. Substorm activity has been previously found to predominantly occur near midnight \citep{Frey2004}. The diurnal pattern of heightened HD activity is consistent with both substorm-driven and impulse-driven GIC. It is expected that the diurnal distribution of HD will be somewhat broader in MLT than that in magnetic data from the AUTUMNX meridian alone due to the longitudinal extent of the power network substations whose HD levels were monitored. We now directly compare HD levels occurring around these impulsive events with periods of similar global electrojet activity, as indicated by electrojet indices.

\subsection{Superposed Epoch Analysis}

We now look at the geoeffectiveness of the impulsive vertical events. Some events in Figure \ref{todplots} appear to be substorm associated, and we would like to minimize the number of this type of event in our comparison analysis. Intensification of the westward auroral electrojet will result in a positive vertical magnetic component perturbation detectable to the north, and a negative perturbation to the south \citep{Kisabeth1974}. For an intensified westward electrojet one would expect a downward perturbation to the vertical component to the north, and upward (negative $B_z$) to the south. This tendency can be seen in the increased number of positive $B_z$ perturbations in the pre-midnight sector at Salluit, and negative $B_z$ perturbations in the pre-midnight sector at Kuujjuarapik. Inukjuak's auroral latitude means that it is more often closer to the latitude of the electrojet. As a result, it should experience less frequent latitudinally offset electrojet enhancements which would produce $B_z$ perturbations than sites to its north or south. Without introducing additional selection criteria, using Inukjuak geomagnetic field data will minimize the number of substorm-related events we use in our analysis. Further, to select for unusually large impulses we have taken the fifty largest negative $B_z$ impulsive events at Inukjuak.

\begin{figure}[ht]
\centering
\includegraphics[width=30pc]{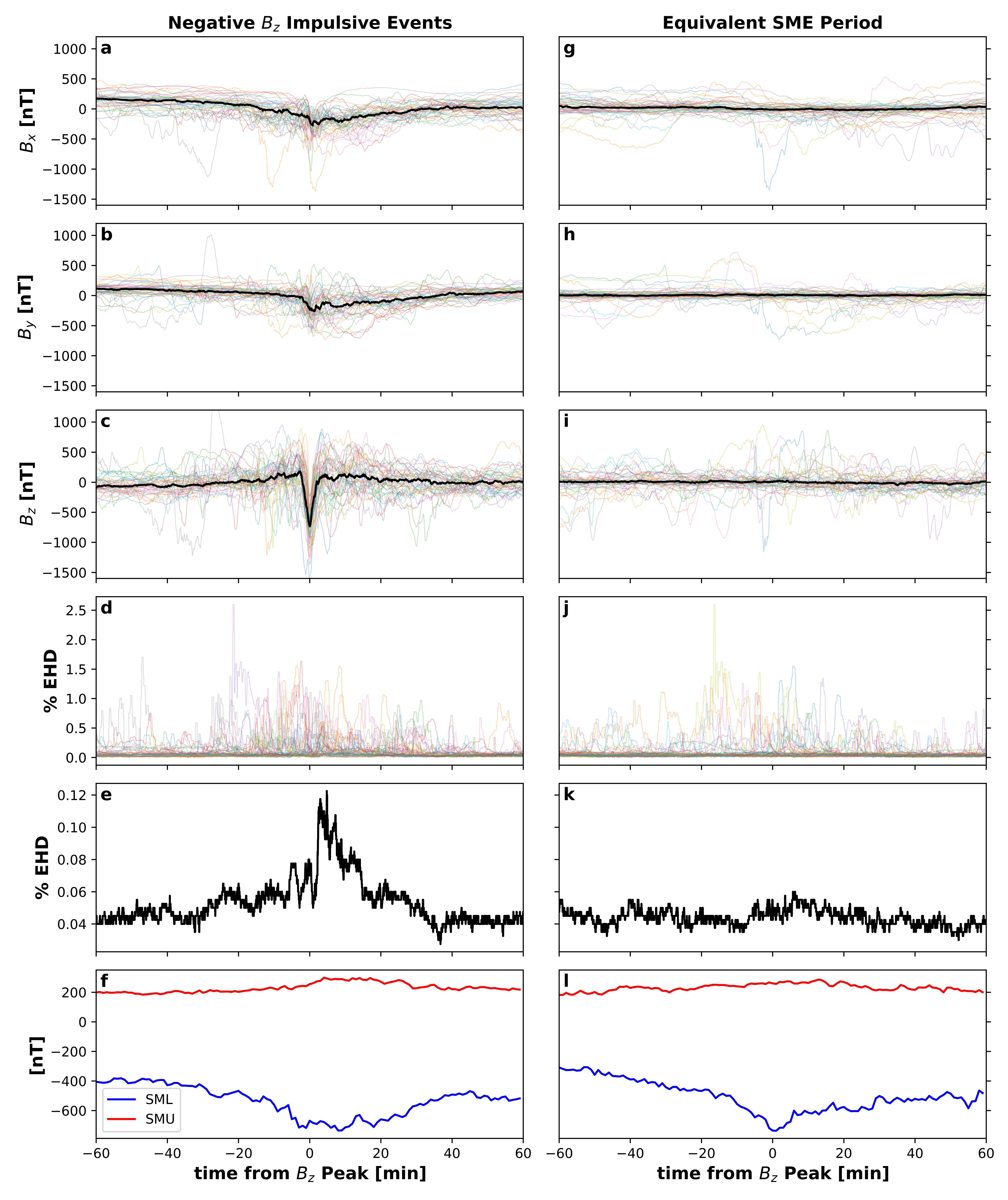}
\caption{Left (a-f): Superposed epoch analysis of magnetometer data from Inukjuak QC, and EHD from the Tilly substation in the Hydro-Qu\'ebec network (median values in bold black). Superposed epoch trigger is the negative $B_z$ extremum for the 50 largest negative $Bz$ impulsive events. Median SMU (red) and SML (blue) for these 2-hour windows are shown in the bottom panel. Individual EHD traces are colorized in \textbf{d}, median values shown in \textbf{e}. Right (g-l): Inukjuak QC magnetometer readings for equivalent SME periods at the $B_z$ extrema, and corresponding median EHD values at the Tilly substation. All SME values have been minimum EHD subtracted for each two hour period. Events without HD data from Tilly were not used (5 of 50 events).}
\label{superposed}
\end{figure}

Superposed epoch analysis for two hour windows around the negative $B_z$ events selected can be seen in Figure \ref{superposed}. Median component values in bold black in panels a-c for separate geomagnetic components demonstrate some average variation in the horizontal geomagnetic field (a,b), but vertical perturbations (c) are much larger. Considerable individual event variability in all components can also be seen.

Averaged EHD data from the Tilly substation are found in panels e and k of Figure \ref{superposed}. This EHD data is taken from the continuous EHD record, rather than the event-limited HD data used for Figure \ref{SMDA-start}. Median EHD response (e) indicates increased geoeffectiveness of negative $B_z$ impulsive events, though individual events (d) do evidence variability in response. SuperMAG auroral indices SML and SMU (analogous to AL and AU) \citep{SuperMAG,SMELi} indicate some strengthening of the westward electrojet prior to the impulsive events, corresponding with a slight median decrease in the SML index. Panels g-l of Figure \ref{superposed} shows data from randomly selected periods with SME values within 20 nT of the SME level at the impulsive event, for the same magnetometer and substation transformer. No median deviations in EHD response are seen for periods with equivalent SME in panel k. This provides evidence for the locality of these vertical impulsive events. Electrojet indices reflect the development of substorm phenomena, but do not well capture the magnitude of local geomagnetic variability associated with these impulsive events.

\section{Discussion}

We have presented evidence of regularly occurring auroral-zone impulsive events. These are principally, though not solely, vertical in nature, and almost entirely found on the nightside. They appear to be localized in latitudinal occurrence and are principally  negative $B_z$ perturbations. The locality and latitudes of occurrence indicate that this is an ionospheric auroral electrojet-related phenomenon. The vertical nature of the disturbances is evidence of vortical or anti-parallel directed ionospheric currents, which are also short-lived in duration, due to the impulsive nature of the $B_z$ perturbations. \citet{doi:10.1029/GL013i011p01089} determined a possible vortical Hall current system driven by field-aligned currents (FAC) that would produce a vertical geomagnetic perturbation. The magnitude of the impulsive events in the present study are larger, with most impulsive $B_z$ events occurring at auroral zone latitudes, but the proposed ionospheric current system is consistent with our observations. Traveling convection vortices (TCV) have been previously observed in ionosphere \citep{FriisChristensen1988,Engebretson2013}. Evidence has also been found of TCVs which are geoeffective \citep{Belakhovsky2019,Chinkin2021}. However, TCVs occur near the dayside with stronger periodicities than we select for. Some $B_z$ impulsive events seen in Figure \ref{todplots}, particularly those in Salluit and Radisson, are likely signatures of substorm electrojet enhancement \citep{Kisabeth1974} rather than ionospheric vorticity, though this is less true of the largest $B_z$ impulses seen at Inukjuak.

It may be surmised through both the auroral latitude and pre-midnight peak occurrence of these impulsive negative $B_z$ events that they could be associated with the Harang reversal \citep{Kamide1978}. This location, where equatorward of the eastward electrojet the westward electrojet can be found \citep{Harang} provides a potential source of current vorticity. However, the largest $B_z$ impulses are both larger in amplitude and shorter in duration than previous ground-based magnetometer observations of the Harang reversal \citep{Kamide1983,Untiedt1993}. An impulsive intensification of eastward and westward convection electrojets could provide a mechanism to intensify the transient $B_z$ signature associated with the Harang reversal. While reduction of SML (and enhancement of the westward electrojet) is evident in panel f of Figure \ref{superposed}, the enhancement of the eastward electrojet as indicated by an increase in SMU is relatively small. Counter-clockwise vortices have also been previously associated with strengthening Region 2 upward FAC in the proximity of the Harang reversal, through combined Poker Flat Incoherent Scatter Radar (PFISR) and THEMIS GBO all-sky imager (ASI) observations of substorm events \citep{Zou2009}.

It is unsurprising that GIC-driven HD events in high latitude sub-auroral power transmission infrastructure would predominantly occur on the nightside around midnight, proximate to the peak of substorm occurrence. However, we have additionally found evidence of median increases in nearby substation transformer HD levels during vertical impulsive events, even when compared with periods with similar SME. The geographic distance between the substation and magnetometer suggests that these impulsive events are localized phenomena, and that their resultant geoelectric fields may be distributed on a scale of 100s of kilometers.

That these impulsive events have ground magnetic signatures which are predominantly vertical in nature, regularly occur, exhibit similar diurnal variations to those of geomagnetically-driven HD events, and are found to excite larger median harmonic distortion responses from nearby substation transformers than periods with equivalent electrojet indices. This indicates that these are events of operational interest to power transmission infrastructure operators at auroral and high-latitude sub-auroral latitudes. Additionally, geoelectric field models which neglect vertical geomagnetic perturbations may systematically under-represent the magnitude of GIC which such impulsive events may produce.

\section{Conclusions}

We present evidence that there are regularly occurring vertical impulsive events in the geomagnetic field at auroral zone latitudes. These events predominantly occur on the nightside in the pre-midnight sector. They exist for minutes in duration and are large in magnitude, with many vertical component perturbations in excess of 1000 nT. A possible driving mechanism of these events is transient ionospheric current vorticity associated with the auroral electrojet. Superposed epoch analysis suggest they occur coincidentally with heightened levels of HD in high-latitude sub-auroral substation transformers, and so present a potential source of space weather disruption in high latitude sub-auroral and auroral zone ground-based infrastructure. The demonstrated presence and geoeffectiveness of these vertical impulsive events presents a space weather hazard to auroral-zone and high-latitude sub-auroral power transmission networks. This emphasizes the need for improved three-dimensional geoelectric field models, as the extent of the space weather hazard posed by vertical impulsive geomagnetic events may not be captured by geoelectric field models which neglect the vertical component of the geomagnetic field.

\begin{acknowledgements}
For the SML and SMU indices, we gratefully acknowledge the SuperMAG collaborators 
(https://supermag.jhuapl.edu/info/?page=acknowledgement). SuperMAG data sets are available from their repository (https://supermag.jhuapl.edu/). AUTUMNX is supported by CSA funding (http://autumn.athabascau.ca), with open data access to magnetic data used in this study. We thank Ian Schofield for his role in installation and operation of the AUTUMNX array. Harmonic distortion data was provided to us by Hydro-Qu\'ebec. Archiving historical GIC and HD data is a time-consuming and costly endeavor for which we would like to thank Hydro-Qu\'ebec, and is a practice we would encourage other utilities to consider emulating.
\end{acknowledgements}


\bibliography{jswsc}
   

\end{document}